\documentclass[aps,prb,reprint,superscriptaddress]{revtex4-2}

\usepackage{amsmath}
\usepackage{amssymb}
\usepackage{xcolor}

\usepackage{amsmath,amssymb,amsfonts,bm,color,graphicx}

\usepackage{tabularx,multirow,array,diagbox}

\usepackage{bm}

\usepackage[unicode=true,colorlinks=true]{hyperref}

\usepackage[etex=true,export]{adjustbox}

\usepackage{ulem}
\renewcommand{\v}[1]{{\bf #1}}

\newcommand{\beq}{\begin{equation}}
\newcommand{\eeq}{\end{equation}}
\newcommand{\beqn}{\begin{eqnarray}}
\newcommand{\eeqn}{\end{eqnarray}}

\usepackage{graphicx} 

\begin{document}

\title{$\phi_0$-junction and Josephson diode effect in high-temperature superconductor}
\author{Guo-Liang Guo}
	\affiliation{School of Physics and Wuhan National High Magnetic Field Center, Huazhong University of Science and Technology, Wuhan, Hubei 430074, China}
 
\author{Xiao-Hong Pan}
	\affiliation{School of Physics and Wuhan National High Magnetic Field Center, Huazhong University of Science and Technology, Wuhan, Hubei 430074, China}
	\affiliation{Institute for Quantum Science and Engineering and Hubei Key Laboratory of Gravitation and Quantum Physics, Wuhan, Hubei 430074, China}
	\author{Xin Liu}
	\email{phyliuxin@hust.edu.cn}
	\affiliation{School of Physics and Wuhan National High Magnetic Field Center, Huazhong University of Science and Technology, Wuhan, Hubei 430074, China}
	\affiliation{Institute for Quantum Science and Engineering and Hubei Key Laboratory of Gravitation and Quantum Physics, Wuhan, Hubei 430074, China}

\begin{abstract}
Motivated by recent progress in both the Josephson diode effect (JDE) and the high-temperature Josephson junction, we propose to realize the JDE in an s-wave/d-wave/s-wave (s-d-s) superconductor junction and investigate the high-temperature superconducting order parameters. The interlayer coupling between s-wave and d-wave superconductors can induce an effective $d+is$ superconducting state, spontaneously breaking time-reversal symmetry. The asymmetric s-d interlayer couplings break the inversion symmetry. Remarkably, the breaking of these two symmetries leads to a $\phi_0$-junction but does not generate JDE. We find that the emergence of the JDE in this junction depends on the $C_4$ rotational symmetry of the system. Although breaking $C_4$ rotational symmetry does not affect time-reversal and inversion symmetries, it can control the magnitude and polarity of diode efficiency. Furthermore, we propose observing C$_{4}$ symmetry breaking controlled JDE through asymmetric Shapiro steps. Our work suggests a JDE mechanism that relies on high-temperature d-wave pairing, which could inversely contribute to a potential experimental method for detecting the unconventional pairing symmetry in superconductors.

\end{abstract}
\maketitle

\section{introduction}
The superconducting diode effect, analogous to the conventional semiconductor diode effect in p-n junction \cite{Scaff1947,Shockley1949}, refers to the nonreciprocal nature of supercurrent. This phenomenon has attracted significant attention in the past few years due to its potential applications in superconducting electronics \cite{Jiang1994,Braginski2019,Ando2020,Miyasaka2021,Zhang2022,Liang2023,Picoli2023,Satchell2023,Hou2023,Banerjee2024}. In such systems, the critical currents along opposite directions exhibit different magnitudes, leading to the dissipationless flow of current in one direction but resistive in the opposite. This effect is prevalent in Josephson junction devices, where it is termed the Josephson diode effect (JDE) \cite{Chen2018,Baumgartner2022,Davydova2022,Pal2022,Steiner2023,Banerjee2023,Legg2023,Maiani2023}. Breaking time-reversal and inversion symmetries is necessary to achieve non-reciprocal supercurrent \cite{Jiang2022, Nadeem2023}. Interestingly, the different symmetry-breaking mechanisms give rise to distinct JDE mechanisms. A typical form of time-reversal symmetry breaking occurs within the electronic Hamiltonian. Magnetic or exchange fields can break time-reversal symmetry, inducing finite Cooper-pair momentum, thereby provides a JDE mechanism, which has been extensively studied both experimentally and theoretically \cite{Yokoyama2014,Alidoust2021,Yuan2022,Legg2022,Jeon2022,Lu2023}. The polarity of the diode efficiency ($\eta$) is usually relate to the direction of the magnetic field or exchange fields. Recent experiments observing JDE at zero magnetic fields suggest new JDE mechanisms based on spontaneous time-reversal symmetry breaking \cite{Golod2022,Narita2022}, such as JDE systems utilizing twisted bilayer or trilayer graphene \cite{Lin2022,Bauriedl2022,DiezMerida2023}, due to the valley polarization and the trigonal warping of the Fermi surface \cite{Hu2023}. JDE at zero magnetic fields has also been reported in recent experiments involving transition metal dichalcogenide (TMD) Josephson junctions \cite{Wu2022}. Beyond the JDE exhibiting time-reversal symmetry breaking within electronic Hamiltonians, the JDE has been experimentally observed in twisted nodal superconductors where spontaneous time-reversal symmetry breaking occurs within the pairing function \cite{Zhu2023,Ghosh2024}. The diode efficiency $\eta$ in such system exhibits a strong dependence on the twist angle $\theta$ \cite{Tummuru2022,Zhao2023,Volkov2024}. Remarkably, even at $\theta=0$, the JDE behavior has been detected without an apparent source of time-reversal breaking \cite{Zhu2023}. This seems to conflict with the expectation of JDE in d-wave pairing Josephson junction \cite{Zhu2021,Zhao2023,Zhu2023}. Therefore, fully understanding JDE in high-$T_{\text{c}}$ superconductor may help to elucidate its pairing function.


\begin{figure}[!htbp]
	\centering
	\includegraphics[width=1\columnwidth]{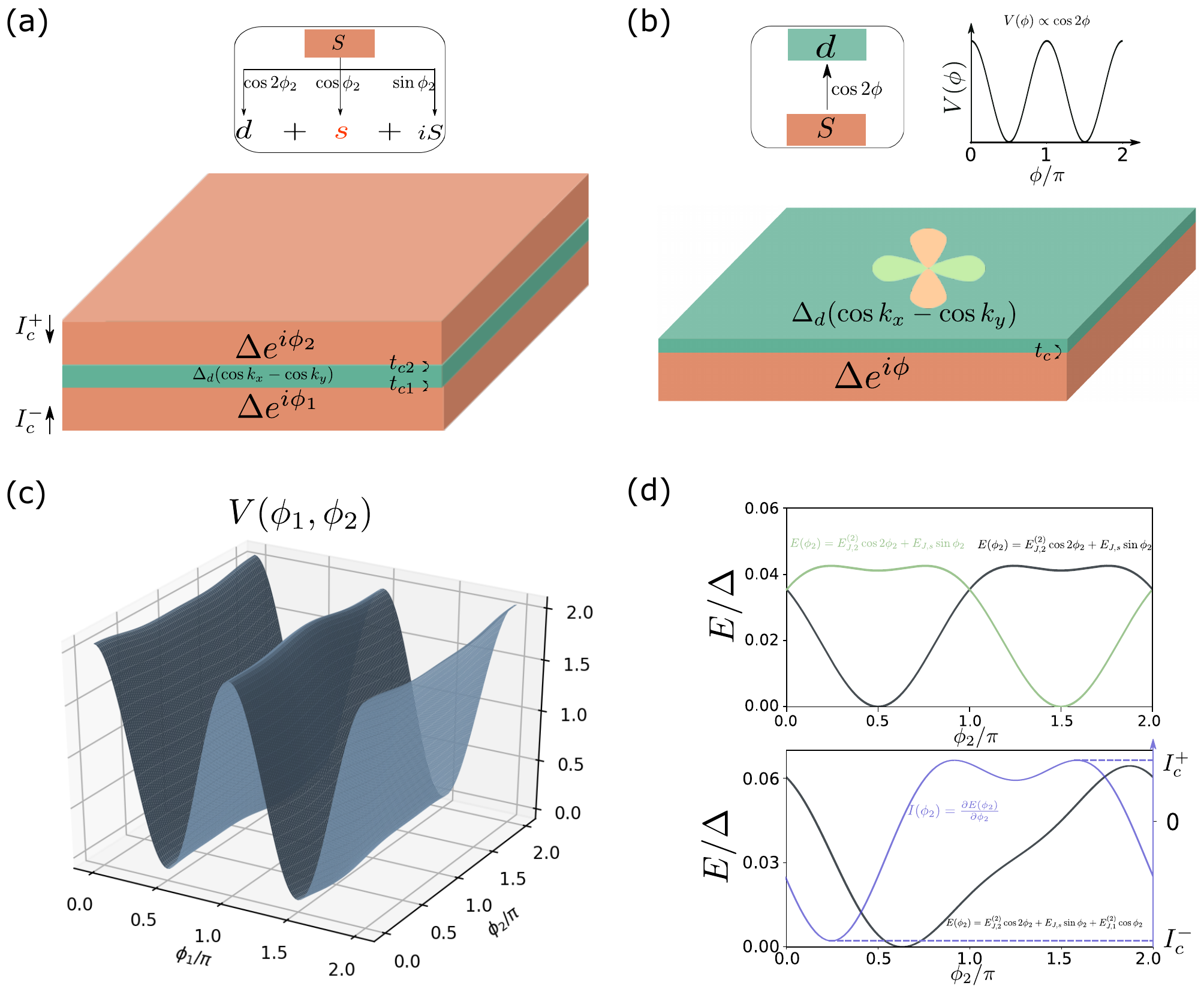}
	\caption{(a) Schematic diagram of the s-d-s junction, which consists of two s-wave and thin d-wave superconductors with asymmetric interlayer couplings $t_{c1}>t_{c2}$. The upper panel shows the origin of the three terms in the free energy, with the red 's' representing the single Cooper pair tunneling ($\cos\phi_2$) induced by the $C_4$ symmetry breaking. (b) Schematic diagram of the s-d junction. The upper panel shows the energy phase relation $V(\phi)\propto\cos2\phi$ of the s-d junction with $C_4$ rotation symmetry. (c) The 2D free energy of s-d-s junction system, which is deeper in $\phi_1$ direction but shallower in $\phi_2$ direction. (d) The energy-phase relation $E(\phi_2)$ of the junction with(without) $C_4$ symmetry. The upper panel, $E(\phi_2+\pi/2)=E(-\phi_2+\pi/2)$ with $\phi_1=\pm \pi/2$, no JDE. The lower panel shows the potential shape at maximum $\eta$ (take $\phi_1=\pi/2$) and the corresponding current-phase relation $I(\phi_2)$.}
\label{fig1}
\end{figure}

In this work, we investigate the JDE in an s-wave/d-wave/s-wave (s-d-s) Josephson junction, as depicted in Fig.~\ref{fig1}(a). Achieving the JDE requires breaking both inversion and time-reversal symmetry. To fulfill these requirements, the system is designed with different interface couplings between the d-wave and the two s-wave superconductors, which breaks the inversion symmetry. The frustration between the s-wave and d-wave pairings induces an effective $d+is$ superconducting state, thereby spontaneously breaking the time-reversal symmetry. Interestingly, while the system becomes a $\phi_0$ junction and exhibits a spontaneous Josephson current (SJc) at $\phi=0$, the JDE does not emerge merely from the breaking of time-reversal and inversion symmetries alone. The system remains symmetric energy-phase relationship, say $E(\phi_i-\phi)=E(\phi_i+\phi)$ for certain value of $\phi_i$. Remarkably, this symmetry in Josephson current-phase relation can be violated through $C_4$ symmetry breaking in either the electronic band structure or the superconducting pairing function. This imply that our result can serve as a potential experimental probe for detecting the pairing function in high-temperature superconductors. The polarity of the JDE, which typically depends on the magnetic field direction, can be controlled by the configuration of $C_{4}$ symmetry breaking in the system. It is important to note that breaking $C_{4}$ symmetry alone does not break either time-reversal or inversion symmetry; our findings provide a novel method for controlling the JDE and detecting the pairing function in c-axis high-$T_{\text{c}}$ superconductor.

\section{Results}
Our proposed s-d-s junction system (Fig.~\ref{fig1}(a)) comprises two s-wave layers and one thin d-wave layer superconductors with asymmetric s-d interlayer coupling constants $t_{1}>t_{2}$. In this case, the system maintains $C_4$ symmetry. The energy-phase relation of s-d junction system is characterized by a $\cos(2\phi)$ form with a positive coefficient \cite{Patel2024}, where $\phi$ denotes the phase difference of the two superconductors (Fig.~\ref{fig1}(b)). Accordingly, the Josephson potential of the s-d-s junction system takes
\begin{equation}
V(\phi_1,\phi_2)=E_{J,2}^{(1)}\cos2\phi_1+E_{J,2}^{(2)}\cos2\phi_2+E_{J,s}\cos(\phi_1-\phi_2),
\label{2d-pot}
\end{equation}
where $\phi_{1(2)}$ the phase difference of the two s-wave superconductors with respect to the d-wave superconductor. The first two terms represent the couplings between the d-wave and the two s-wave superconductors, describing the co-tunneling of two Cooper pairs, simultaneously. The corresponding coefficients, $E_{J,2}^{(1)}$ and $E_{J,2}^{(2)}$, are positive and proportional to the interlayer couplings $t_{1}^4$ and $t_{2}^4$ \cite{Patel2024,supp} in the weak coupling regime, respectively. The final term describes the Cooper pair tunneling between the two s-wave superconductors, with the coefficient proportional to $t_{1}^2t_{2}^2$, as well as controlled by the thickness of d-wave superconductors. Here, the subscript 1,2 indicates the order of the cosine term, and the superscript (1), (2) correspond to $\phi_1$ and $\phi_2$, respectively.

\subsection{$\phi_0$ junction with $C_4$ symmetry}
In the s-d-s junction system, to break inversion symmetry, we anticipate the interlayer coupling strength $t_{1}$ to be larger than $t_{2}$, a condition that can be experimentally achieved by modulating the interface couplings \cite{Yang2018}. These asymmetric interlayer couplings lead to a two-dimensional(2D) potential (Eq.~\eqref{2d-pot}) profile that is deeper for $\phi_1$ and shallower for $\phi_2$ (Fig.~\ref{fig1}(c)). This asymmetry is reflected in the relative magnitudes of the coefficients of the 2D potential (Eq.~\eqref{2d-pot}), $E_{J,2}^{(1)}>>E_{J,2}^{(2)}, E_{J,s}$. Consequently, the wavefunction distribution for the lowest two energy states of the potential is narrow in $\phi_1$ but wider in $\phi_2$ \cite{supp}. The phase difference $\phi_1$ is thus locked to one of its minimal points, $\phi_1=\pm\pi/2$, which spontaneously breaks the time-reversal symmetry of the system \cite{Patel2024,supp}. The s-d-s junction can thus be viewed as the effective coupling of s-wave and $d\pm is$-wave superconductors. The 2D potential simplifies to a one-dimensional (1D) form $V(\phi_1=\pm\pi/2,\phi_2)$, denoted hereafter as
\begin{equation}
    E(\phi_2)=E_{J,2}^{(2)}\cos2\phi_2\pm E_{J,s}\sin\phi_2,
    \label{1d-c4}
\end{equation}
where $\pm$ corresponding to $\phi_1=\pm\pi/2$. The potential shape becomes that of a $\phi_0$ junction \cite{Buzdin2003} with the minimal point located at neither 0 nor $\pi$. There exists SJc at $\phi_2=0$ as
\begin{equation}
    I(\phi_2=0)=\frac{\partial E(\phi_2)}{\partial\phi_2}|_{\phi_2=0}=\pm E_{J,s},
\end{equation}
which corresponds to the spontaneous time-reversal symmetry breaking. The direction of SJc is determined by the minimum point of $\phi_1$, which can be adjusted by applying an external current bias that exceeds the maximal current in the $\phi_1$ direction (about $2E_{J,2}^{(1)}$) in experiments \cite{Zhao2023}. 
However, different from the general JDE with inversion and time-reversal symmetries breaking, the JDE does not appears in this junction even both the two symmetries are broken. The energy-phase relation in Eq.~\eqref{1d-c4} retains the symmetry $E(\phi_2+\pi/2)=E(-\phi_2+\pi/2)$ due to the $\pi$ periodicity of $\cos(2\phi_2)$ term (upper panel in Fig.~\ref{fig1}(d)). This indicates that the $\phi_0$ junction is not a sufficient condition for the supercurrent nonreciprocity \cite{Reinhardt2024}.

\subsection{JDE with $C_4$ symmetry breaking}

Nonetheless, the JDE becomes feasible when the system also breaks the $C_4$ rotation symmetry. This symmetry breaking facilitates single Cooper pair tunneling and introduce an $E_{J,1}^{(2)}\cos\phi_2$ term in the free energy as
\begin{equation}
    E(\phi_2)=E_{J,2}^{(2)}\cos2\phi_2\pm E_{J,s}\sin\phi_2+E_{J,1}^{(2)}\cos\phi_2,
    \label{1d-pot}
\end{equation}
where the magnetic and sign of the coefficient $E_{J,1}^{(2)}$ are controlled by the $C_4$ symmetry breaking. With this potential form, the JDE can appear with maximal diode efficiency $\eta=1/3$ \cite{Haenel2022,supp} under the condition $|E_{J,s}|=|E_{J,1}^{(2)}|=2\sqrt{2}E_{J,2}^{(2)}$ (lower panel in Fig.~\ref{fig1}(d)). In this junction, the appearance of JDE is determined by the non-magnetic control of $C_4$ symmetry breaking, corresponding to the presence of the $\cos\phi_2$ term in free energy Eq.~\eqref{1d-pot}. Moreover, given that the coefficient $E_{J,2}^{(2)}$ of $\cos2\phi_2$ term in s-d-s junction is positive \cite{Patel2024,supp}, the polarity of $\eta$ can be reversed by reversing the sign of $E_{J,s}, E_{J,1}^{(2)}$ \cite{supp}, which are influenced by the minimal point of $\phi_1$ and the extent of $C_4$ symmetry breaking in experiment. Note that, the $C_4$ symmetry breaking also introduces $E_{J,1}^{(1)}\cos\phi_1$ term in the potential, which may slightly shift the minimum point of $\phi_1$ from $\pm\pi/2$. However, due to the large interlayer couplings $t_{c1}$, the ratio of the coefficient $E_{J,1}^{(1)}/E_{J,2}^{(1)}$ is significantly smaller than that of $E_{J,1}^{(2)}/E_{J,2}^{(2)}$ for a fixed $C_4$ symmetry breaking term. Therefore, the minimum point of $\phi_1$ remains pinned near $\phi=\pi/2$ and can still be approximated as an effective $d+is$ superconductor. The effective 1D potential $E(\phi_2)$ retains the form shown in Eq.~\eqref{1d-pot} and this conclusion will be substantiated in the following sections.

\section{theoretical and numerical results of s-d-s junction system}

\subsection{theoretical results with perturbation theory}
To investigate the Josephson potential form of this system, we first focus on the energy-phase relation in the s-d junction. The junction comprises conventional s-wave and d-wave superconductors, as depicted in Fig.~\ref{fig1}(b). In the basis $(d_{k,\uparrow},d_{-k,\downarrow}^{\dagger},s_{k,\uparrow},s_{-k,\downarrow}^{\dagger})$, the microscopic Hamiltonian Hamiltonian is
\begin{equation}
\begin{aligned}
\mathcal{H} &=\left(\begin{array}{cc}
h_d & T  \\
T^{\dagger} & h_s 
\end{array}\right), \\
\end{aligned}
\end{equation}
where $h_{d(s)}=\epsilon_k^{d(s)}\tau_z+\Delta_k^{d}(\Delta_s)\tau_x$ describes the isolated d(s)-wave superconductors. Here, $\epsilon_k^{d(s)}$ represents the kinetic energy relative to the Fermi surface, and $\Delta_k^d=\Delta_d(\cos k_x-\cos k_y)$ is the pairing function of the d-wave superconductor, with $\Delta_d$ pairing strength. $\Delta_s=\Delta e^{i\phi}$ the s-wave pairing function with the phase difference $\phi$, $T=t_c\tau_z$ is the coupling term of the s-wave and d-wave superconductors, $t_c=t_0+t_x\cos k_x+t_y\cos k_y$ with $t_0,t_{x(y)}$ the zero-order and first-order of interlayer couplings \cite{supp}. Since the d-wave superconductor features gapless points on the Fermi surface, while the s-wave superconductor is fully gapped, we can perform a Schrieffer-Wolff transformation \cite{Yao2007} in the weak coupling limit ($t_c<\Delta$) to derive an effective d-wave superconductor Hamiltonian and future calculate the free energy of the system as \cite{supp}
\begin{equation}
\begin{aligned}
E_g&=N_F\int_0^{2\pi}d k_x\int_0^{2\pi}d k_y[({\tilde{\epsilon}_k}-\sqrt{{\tilde{\epsilon}}_k^2+D^2})\\
&-\frac{p}{2\sqrt{{\tilde{\epsilon}}_k^2+D^2}}+\frac{p^2}{8({\tilde{\epsilon}}_k^2+D^2)^{3/2}}],
\label{eqfe}
\end{aligned}
\end{equation}
where $D^2=\Delta_d^2(\cos k_x-\cos k_y)^2+m_k^2$, $p=2\Delta_dm_k(\cos k_x-\cos k_y)\cos\phi$, $m_k=t_c^2/\Delta$, $N_F=2\cdot(\frac{L}{2\pi})^2$ the density of state.
The leading contribution to the free energy comes from the $\cos2\phi$ term, corresponding to the co-tunneling of even numbers of Cooper pairs. The associate coefficient $E_{J,2}$ is
\begin{equation}
E_{J,2} = N_F\int_0^{2\pi}d k_x\int_0^{2\pi}d k_y\frac{\Delta_d^2(\cos k_x-\cos k_y)^2m_k^2}{4({\tilde{\epsilon}}_k^2+D^2)^{3/2}}.
\label{coef2}
\end{equation} 
It is evident that the coefficient $E_{J,2}$ of the leading term $\cos(2\phi)$ is positive, not negative, and proportional to interlayer coupling $t_c^4$. Hence, the potential minima of s-d junction occurs at $\phi=\pm\pi/2$ (Fig.~\ref{fig1}(b)) \cite{Patel2024}. With strong interlayer coupling and negligible charge energy, the phase difference fluctuations will vanish and be locked to one of its minimal points, $\phi=\pm\pi/2$. This induces an effective $d\pm is$ superconductor, spontaneously breaking the time-reversal symmetry of the system \cite{Patel2024}.

Notably, the prohibition of the tunneling of an odd number of Cooper pairs, corresponding to the absence of $\cos(2n+1)\phi$ terms in the free energy is contingent upon preserving $C_4$ rotation symmetry in d-wave superconductors. If this symmetry is disrupted by strains, modifying the interlayer coupling to $t_{x(y)}=t_s\pm t_{as}$, with nonzero $t_{s(as)}$ the symmetric(anti-symmetric) interlayer couplings respectively, the tunneling of odd numbers of Cooper pairs becomes permissible, leading to the emergence of $\cos(2n+1)\phi$ terms in the free energy. Specifically, the coefficient $E_{J,1}$ of the leading $\cos(\phi)$ term is expressed as
\begin{equation}
    E_{J,1} = N_F\int_0^{2\pi}d k_x\int_0^{2\pi}d k_y\frac{\frac{2t_0\cdot t_{as}}{\Delta}\Delta_d(\cos k_x-\cos k_y)^2}{({\tilde{\epsilon}}_k^2+D^2)^{3/2}},
\label{coef1}
\end{equation}
which is related to the $C_4$ symmetry breaking term, $t_{as}/t_s$, and interlayer coupling $t_0t_s$. The Josephson potential can subsequently be written as
\begin{equation}
   E(\phi)=E_{J,2}\cos2\phi+E_{J,1}\cos\phi.
\label{toy-pot}
\end{equation}
Due to the $\cos\phi$ term, the minimum of the potential shifts from $\pi/2$ to $\phi_{\rm min}=\pi/2+\delta\phi$, where $\delta \phi=\arcsin(E_{J,1}/4E_{J,2})$. Remarkably, the coefficient of $\cos\phi$ in Eq.~\eqref{coef1} is of second order concerning the coupling strength, corresponding to the tunneling of single Cooper pairs. In contrast, the coefficient of $\cos2\phi$ in Eq.~\eqref{coef2} is of fourth order, corresponding to the co-tunneling of double Cooper pairs. Consequently, with a fixed $C_4$ symmetry breaking term, $t_{as}/t_s$, the ratio of the coefficient $r\equiv E_{J,1}/E_{J,2}$ decreases as the coupling strength increases in the weak coupling regime. Therefore, with fixed $C_4$ symmetry breaking term, strong interlayer coupling hinders shifts in the free energy minimum, which remains pinned near $\phi=\pm\pi/2$, resulting in a $d\pm is$ superconductor.

Though the minimal point $\phi=\pm\pi/2$ break the time-reversal symmetry in the s-d junction, the potential in Eq.~\eqref{toy-pot} retains the time-reversal symmetry, $E(\phi)=E(-\phi)$, and there are no JDE in this junction. Then, we consider the proposed s-d-s junction with a thin d-wave superconductor and asymmetric s-d interlayer couplings ($t_{1}>t_{2}$). There is no direct coupling between the two s-wave superconductors. In the basis $(d_{k,\uparrow},d_{-k,\downarrow}^{\dagger},s_{1,k,\uparrow},s_{1,-k,\downarrow}^{\dagger},s_{2,k,\uparrow},s_{2,-k,\downarrow}^{\dagger})$, the total microscopic Hamiltonian of the system can be expressed as
\begin{equation}
\begin{aligned}
\mathcal{H} &=\left(\begin{array}{ccc}
h_d & T_1 & T_2  \\
T_1^{\dagger} & h_{s1} & 0\\ 
T_2^{\dagger} & 0 & h_{s2}\\ 
\end{array}\right), \\
\end{aligned}
\end{equation}
where $T_{i}=t_{ci}\tau_z$, and $t_{ci}=t_i+t_{ix}\cos k_x+t_{iy}\cos k_y(i=1,2)$, represent the coupling between the s-wave and d-wave superconductors. $h_{d(s_1,s_2)}$ the Hamiltonian of isolated d-wave and the two s-wave superconductors, respectively. By employing the Schrieffer-Wolff transformation, we derive the effective Hamiltonian around the gapless point of the d-wave superconductor and subsequently determine the free energy of the system \cite{supp}.
\begin{equation}
\begin{aligned}
E'_g&=N_F\int_0^{2\pi}d k_x\int_0^{2\pi}d k_y[({\tilde{\epsilon}_k^{'2}}-\sqrt{{\tilde{\epsilon}_k^{'2}+D'^2}})\\
&-\frac{p'}{2\sqrt{\tilde{\epsilon}_k^{'2}+D'^2}}+\frac{p'^{2}}{8(\tilde{\epsilon}_k^{'2}+D'^2)^{3/2}}+O(p'^{3})],
\label{eqfe2}
\end{aligned}
\end{equation}
where $D'^{2}=\Delta_d^2(\cos k_x-\cos k_y)+m_{k1}^2+m_{k2}^2$, and $p'=2m_{k1}\Delta_d(\cos k_x-\cos k_y)\cos\phi_1+2m_{k2}\Delta_d(\cos k_x-\cos k_y)\cos\phi_2+2 m_{k1}m_{k2}\cos(\phi_1-\phi_2)$, with $m_{ki}=t_{ci}^2/\Delta(i=1,2)$.
Obviously, it is a 2D potential $V(\phi_1,\phi_2)$ and there are no $\cos\phi_1(\phi_2)$ terms if the system maintains $C_4$ rotation symmetry (Eq.~\eqref{2d-pot}). The junction, designed with a significant difference in interlayer couplings $t_{c1}>t_{c2}$, results in the 2D potential being deeper along the $\phi_1$ direction but shallower along $\phi_2$ direction (Fig.~\ref{fig1}(c)) \cite{supp}, locking the phase difference $\phi_1$ to one of its minimal points, chosen here as $\phi_1=\pi/2$, thereby spontaneously breaking the time-reversal symmetry of the system. The junction can then be viewed as the effective coupling of $s$-wave and $d+is$-wave superconductors. The potential simplifies to a 1D form $V(\phi_1= \pi/2,\phi_2)$, denoted as $E(\phi_2)$ (Eq.~\eqref{1d-c4}), with the coefficients given by
\begin{equation}
    \begin{aligned}
    E_{J,s}= & N_F\int_0^{2\pi}d k_x\int_0^{2\pi}d k_y\left(-\frac{m_{k1}m_{k2}}{\sqrt{\tilde{\epsilon}_k^{'2}+D'^2}}\right),\\
    E_{J,2}^{(2)}= &N_F\int_0^{2\pi}d k_x\int_0^{2\pi}d k_y\frac{m_{k2}^2\Delta_d^2(\cos k_x-\cos k_y)^2}{2(\tilde{\epsilon}_k^{'2}+D'^2)^{3/2}}.
    \end{aligned}
\label{eq-cof}
\end{equation}
The two terms are proportional to interlayer couplings $t_{c1}^2t_{c2}^2$ and $t_{c2}^4$, respectively. The other terms $\cos(2n+1)\phi_{1(2)}$ vanish if the system maintains $C_4$ rotation symmetry. Thus, the system acts as a $\phi_0$ junction \cite{Liu2016} with a  SJc $I_{SJc}\approx \pm E_{J,s}$ at $\phi_2=0$, depending on $\phi_1=\pm\pi/2$.

However, the presence of only $\cos(2\phi_2)$ and $\sin(\phi_2)$ components in the free energy remains the symmetry $E(\phi_2+\pi/2)=E(-\phi_2+\pi/2)$ (Fig.~\ref{fig1}(d)), prohibits the appearance of the JDE in this case, even if the system breaks both inversion and time-reversal symmetries. The JDE becomes feasible if the $C_4$ rotation symmetry is broken, denoted by $t_{ix(y)}=t_{is}\pm t_{ias}(i=1,2)$, achievable by strain in experiment \cite{supp}. For convenience, we assume the $C_4$ symmetry breaking for the two s-d interlayer couplings takes the same value, $t_{1as}/t_{1s}=t_{2as}/t_{2s}$. This disruption introduces an additional term $E_{J,1}^{(2)}\cos(\phi_2)$ in the free energy Eq.~\eqref{eqfe2}, with the associated coefficient as
\begin{equation}
    E_{J,1}^{(2)} = \pm N_F\int_0^{2\pi}d k_x\int_0^{2\pi}d k_y\frac{\frac{t_2t_{2as}}{\Delta}\Delta_d(\cos k_x-\cos k_y)^2}{\sqrt{\tilde{\epsilon}_k^{'2}+D'^2}},
\label{c1}
\end{equation}
where "$\pm$" correspond to $\pm t_{2as}$, controlled by the strain in the $x$ or $y$ directions in experiments. Additionally, the free energy also includes the term $E_{J,1}^{(1)}\cos\phi_1$ if the $C_4$ symmetry is broken by strains, which may slightly shift the minimum point of $\phi_1$ from $\pi/2$. However, due to the larger interlayer coupling $t_{1}$, the ratio of the coefficient $E_{J,1}^{(1)}/E_{J,2}^{(1)}$ is significantly smaller than that of $E_{J,1}^{(2)}/E_{J,2}^{(2)}$ for a fixed $C_4$ symmetry breaking $t_{as}/t_s$, as corroborated by the calculations in Eq.~\eqref{coef2}, \eqref{coef1}. Therefore, the minimum point of $\phi_1$ remains near $\pi/2$ and can still be approximated as an effective $d+is$-wave superconductor. As a result, with the leading contribution terms, the free energy is formalized as Eq.~\eqref{1d-pot}, predominantly influenced by the interlayer coupling strength $t_{2}$, the thickness of d-wave superconductor and the $C_4$ symmetry breaking term $t_{as}/t_s$ in the junction. Note that the slightly shift of the minimum point of $\phi_1$ from $\pi/2$ merely modifies the coefficients and does not alter the form of Eq.~\eqref{1d-pot} \cite{supp}. With this unconventional free energy, the JDE can manifest, and the polarity of $\eta$ can be reversed by changing the sign of $E_{J,1}^{(2)}$ or $E_{J,s}$, which is determined by the sign of $C_4$ symmetry breaking strength (Eq.~\eqref{c1}) and the minimal point of $\phi_1$ \cite{supp}. 

\subsection{numerical results}
The theoretical calculations are valid in the weak coupling limit ($t_c<\Delta$). To extend this analysis, we perform numerical simulations of the s-d junction system by constructing a lattice model for the junction \cite{supp} beyond the weak coupling limit. The numerical simulations are conducted using the Kwant program \cite{Groth2014}. In the Nambu space, the TB Hamiltonian of the system is given by
\begin{equation}
\begin{aligned}
h_s = & \epsilon_k^s\tau_z+\Delta(\cos\phi\tau_x+\sin\phi\tau_y),\\
h_d = & \epsilon_k^d\tau_z+\Delta_k^d\tau_x,
\label{tbh}
\end{aligned}
\end{equation}
with $h_{s(d)}$ represents the isolated Hamiltonian of s(d)-wave superconductor. $\epsilon_k^s=2t(3-\cos k_x-\cos k_y-\cos k_z)-\mu_s$ the kinetic energy of the s-wave superconductor, with $t$ the isotropic hopping strength in three directions, and $\Delta$ is the pairing strength of s-wave superconductor. For the d-wave layer, $\epsilon_k^d=2t(2-\cos k_x-\cos k_y)+2t_{dz}(1-\cos k_z)-\mu_d$, where $t$ ($t_{\textrm{dz}}$) is the in-plane (out-of-plane) hopping, and usually $t_{dz}<t$ \cite{Liao2018,Zhu2023}. $\mu_{s(d)}$ the chemical potential of the s(d)-wave superconductor. $\Delta_k^d=\Delta_{d}(\cos k_x-\cos k_y)$ the pairing function of the d-wave superconductor. The interlayer coupling is described by $h_\textrm{t}=t_c\tau_z$, parameterized by the interface hopping strength $t_c=t_0+t_x\cos k_y+t_y\cos k_y$, $t_{x(y)}=t_s\pm t_{as}$.

\begin{figure}[!htbp]
	\centering
	\includegraphics[width=1\columnwidth]{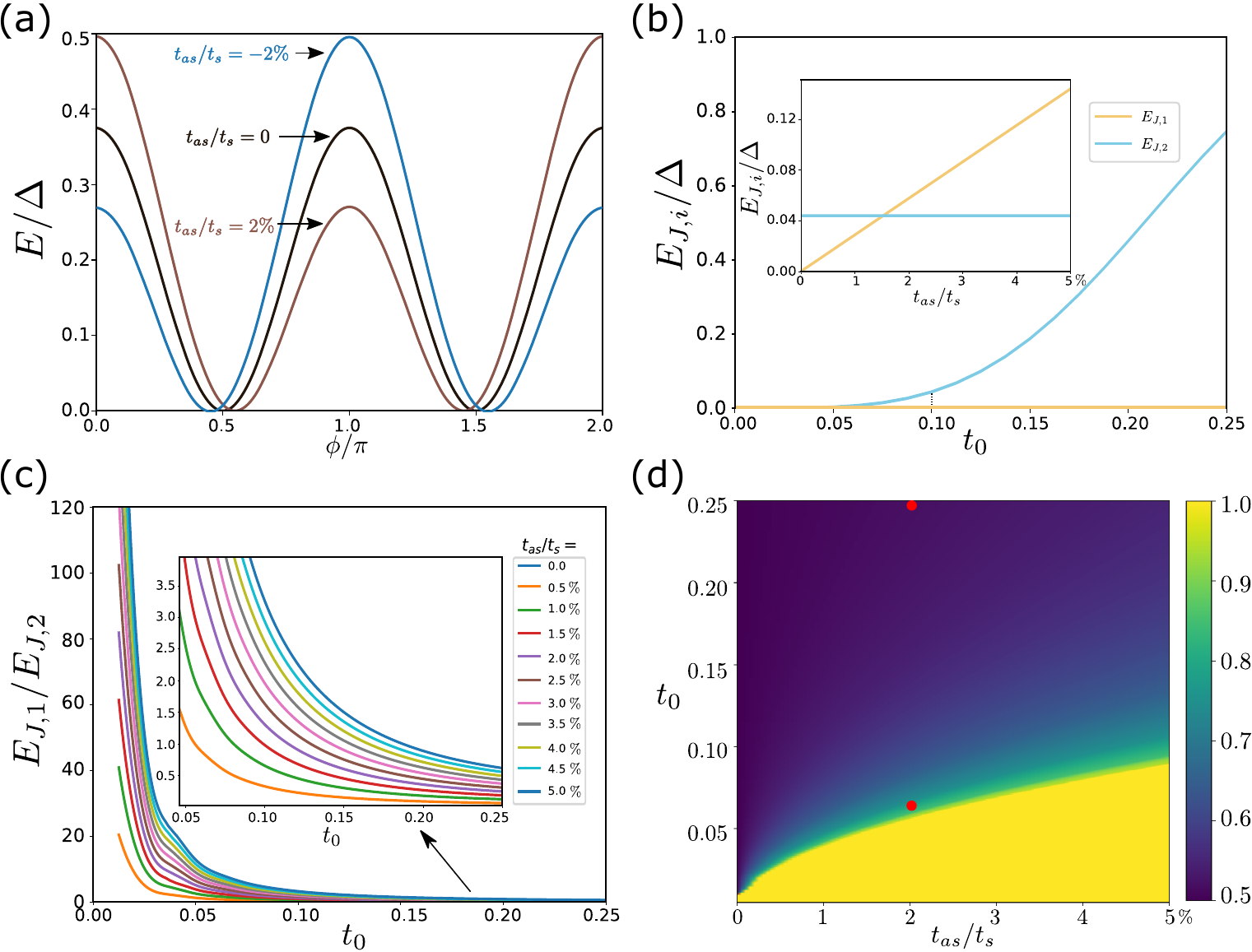}
	\caption{(a) Free energy of the s-d junction system with $t_0=0.15$. The three lines correspond to $t_{as}/t_s=0,\pm 2\%$. (b) The coefficients $E_{J,1}$ and $E_{J,2}$ changes with coupling strength $t_0$. Inset shows the coefficient changes with $C_4$ symmetry breaking $t_{as}/t_s$ with fixed $t_0=0.1$, corresponding to the dashed vertical black line. (c) The ratio of coefficient $E_{J,1}/E_{J,2}$ changes with coupling strength $t_0$ for different symmetry breaking strength $t_{as}/t_s$. (d) The minimal point $\phi_{\rm min}/\pi$ of the free energy changes with $t_{as}/t_s$ and interlayer coupling strength $t_0$. The red dots are the parameters used in the s-d-s junction simulation. The parameters are $t=1, t_{dz}=0.15$, $\mu_s=\mu_d=2.5$, $\Delta=0.05,\Delta_{d}=0.2$, $L_d=5,L_s=30$.}
\label{fig2}
\end{figure}

The system maintains translation symmetry in $x$ and $y$ directions. We thus apply periodic boundary conditions along these directions. The Andreev levels of the system are deduced by evaluating the eigenenergies of the TB Hamiltonian as a function of the phase difference $\phi$. Then, the zero temperature free energy $F(\phi)$ of the junction is then calculated by summing all negative eigenenergies of the Andreev levels and performing summations over $k_x,k_y$ (discretized at 30 equidistant points ranging from 0 to $2\pi$). The results, shown by the black line in Fig.~\ref{fig2}(a), reveal that the Josephson potential exhibits a $\pi$ periodicity and is primarily dominated by a $\cos2\phi$ term with a positive coefficient if the system maintains $C_4$ rotation symmetry, $t_{as}=0$. The corresponding minimal point of the potential is $\phi=\pm\pi/2$. However, the violation of $C_4$ rotation symmetry breaking, characterized by non-zero term $t_{as}$ paves the way for single Cooper pair tunneling, introducing $\cos\phi$ term in the potential. The sign of the coefficient $\cos\phi$ is relevant to $t_{as}/t_s$, as depicted by the blue and brown lines in Fig.~\ref{fig2}(a). This observation is consistent with the theoretical calculations (Eq.~\eqref{coef1}). With the Josephson potential, we can do Fourier transformation to get the coefficients $E_{J,2}$ and $E_{J,1}$ for the leading $\cos2\phi$ and $\cos\phi$ terms, respectively. The coefficient $E_{J,2}$ increase with the coupling strength $t_0$, whereas $E_{J,1}$ remains null if the $C_4$ rotation symmetry is maintained, as shown in Fig.~\ref{fig2}(b). Upon breaking $C_4$ symmetry, $E_{J,1}$ emerges and is proportional to the $t_{as}/t_s$, while $E_{J,2}$ experiences only minor changes, as illustrated in the inset of Fig.~\ref{fig2}(b). Future, we calculate the ratio of the coefficient $E_{J,1}/E_{J,2}$ in the presence of $C_4$ symmetry breaking. It decreases as the coupling strength $t_0$ increases with a fixed $C_4$ symmetry breaking strength $t_{as}$, as shown in Fig.~\ref{fig2}(c). Consequently, with the Josephson form in Eq.~\eqref{toy-pot}, under a fixed $C_4$ symmetry breaking term $t_{as}/t_s$, the minimal point of the potential decreasingly shifts with larger interface coupling strength, and the minimal point of the potential remains in the vicinity of $\pm\pi/2$, as shown in Fig.~\ref{fig2}(d), and still results in a $d+is$ superconductor. The results are consistent with the theoretical calculations.

\begin{figure}[!htbp]
	\centering
	\includegraphics[width=1\columnwidth]{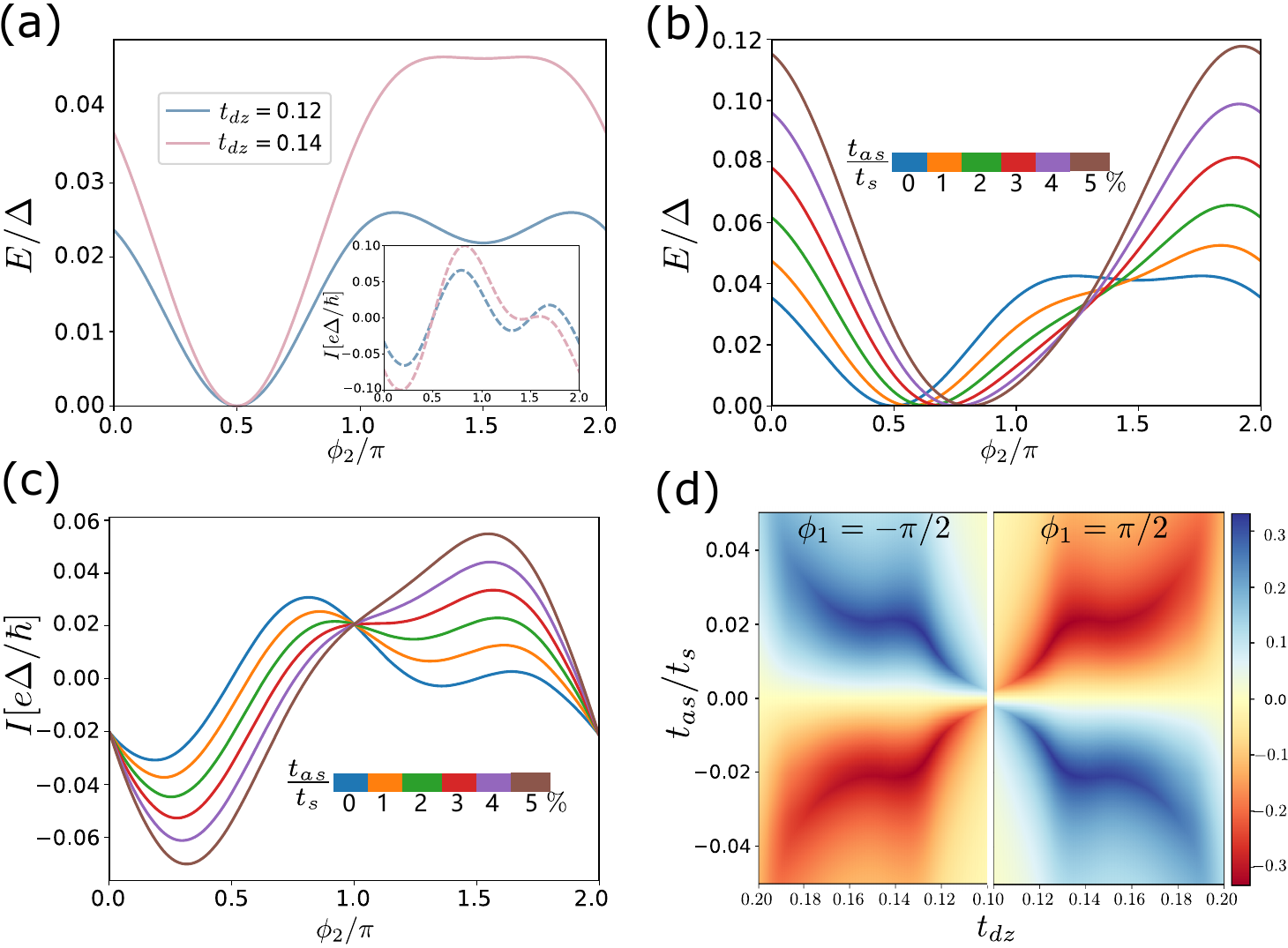}
	\caption{(a) Free energy of s-d-s junction for different values of $t_{dz}$, inset is the corresponding energy-phase relation. (b) Free energy of the s-d-s junction system for different $t_{as}/t_s$ at $t_{dz}=0.13$, corresponding to the vertical dashed lines in (d). (c) The current phase relation for different ratios of $t_{as}/t_s$. (d) Diode efficiency changes with $t_{as}/t_s$ and $t_{dz}$ for $\phi_1=\pm\pi/2$. In the numerical calculating, the parameters are $t=1$, $\mu_s=\mu_d=2.5$ $t_1=0.25,t_2=0.06$(red dots in Fig.~\ref{fig2}(d)), $\Delta=0.05,\Delta_d=0.2$, $L_s=30,L_d=5$.
  }
\label{fig3}
\end{figure}

Then, we conduct numerical simulations of the s-d-s junction system using a lattice model \cite{supp}. In this setup, we assume that one of the s-d interface couplings is notably stronger, locking the corresponding phase to one of its minima. Here, we take $t_{1}>t_{2}$ and lock the phase $\phi_1$ to its minimum of $\pi/2$ \cite{supp}, thereby spontaneously breaking the time-reversal symmetry. The potential of the system can be calculated and it is primarily dominated by  $E_{J,2}^{(2)}\cos2\phi_2$ and $E_{Js}\sin\phi_2$ terms (Fig.~\ref{fig3}(a)). Increasing $t_{dz}$, which is equivalent to tuning the thickness of d-wave superconductors experimentally, enhances the $\sin\phi_2$ component in the free energy as it increases the effective coupling of the two s-wave superconductors (Fig.~\ref{fig3}(a)). There exists a spontaneous Josephson current at $\phi_2=0$ (inset in Fig.~\ref{fig3}(a)), corresponding to the spontaneous time-reversal symmetry breaking. However, the JDE can not appear in this scenario as the energy-phase relation remains the symmetry $E(\phi_2+\pi/2)=E(-\phi_2+\pi/2)$ (Fig.~\ref{fig3}(a))

To investigate the JDE in this junction, we next introduce the $C_4$ symmetry breaking in electronic structure by setting $t_{ix(y)}=t_{is}\pm t_{ias}$ with $t_{as}/t_{s}$ describe the $C_4$ symmetry breaking. This introduces an additional $E_{J,1}^{(2)}\cos\phi_2$ term in the potential as calculated in Eq.~\eqref{c1} and reshapes the energy-phase relation as shown in Fig.~\ref{fig3}(b). Consequently, the current phase relation is also altered, as illustrated in Fig.~\ref{fig3}(c), clearly demonstrating asymmetry in the critical currents in opposite directions. Finally, the diode efficiency $\eta$ of the system versus $t_{dz}$ and $t_{as}/t_s$ is calculated in Fig.~\ref{fig3}(d), showing a maximum diode efficiency of up to $\frac{1}{3}$. This also indicates that the polarity of the diode efficiency can be reversed by the opposite $C_4$ symmetry breaking strength and minimal point of $\phi_1$. Critically, it clearly shows that there is no JDE ($\eta=0$) if the junction maintains $C_4$ symmetry ($t_{as}=0$).

Note that in both our theoretical and numerical calculations of s-d-s junction, the free energy $F(\phi_2)$ is obtained by fixing the phase $\phi_1=\pm\pi/2$, which is reasonable in the limit $t_{1}>t_{2}$. The breaking of $C_4$ symmetry will also generate a $\cos\phi_1$ term in the system, potentially shifting the minimum point of $\phi_1$ from $\pi/2$. However, as indicated by our numerical calculations, the maximum diode efficiency is achieved with the ratio of the coefficient $E_{J,1}^{(2)}/E_{J,2}^{(2)}=2\sqrt{2}$. Given the substantial difference in interlayer coupling strengths $t_{1}>t_{2}$, and referring to Fig.~\ref{fig2}(c),(d) and Eq.~\eqref{toy-pot}, we estimate that the shift in the minimum point of $\phi_1$ from $\pi/2$ is on the order of $10^{-2}\pi$. We also assume that the $C_4$ symmetry breaking strength $t_{ias}/t_{is}$ is consistent for both interlayer couplings. This approximation can not change the conclusion that the shift in the minimum point of $\phi_1$ from $\pi/2$ is small, as long as the coupling strength $t_{1}$ is strong. Therefore, the phase $\phi_1$ can still be approximated as $\pi/2$, and this approximation does not alter the conclusions drawn. In experimental settings, with thin d-wave layers and asymmetric interlayer coupling strengths, the JDE is expected to manifest as long as the $C_4$ rotation symmetry is broken within the d-wave superconductor.

\section{discussion}
\begin{figure}[!htbp]
	\centering
	\includegraphics[width=1\columnwidth]{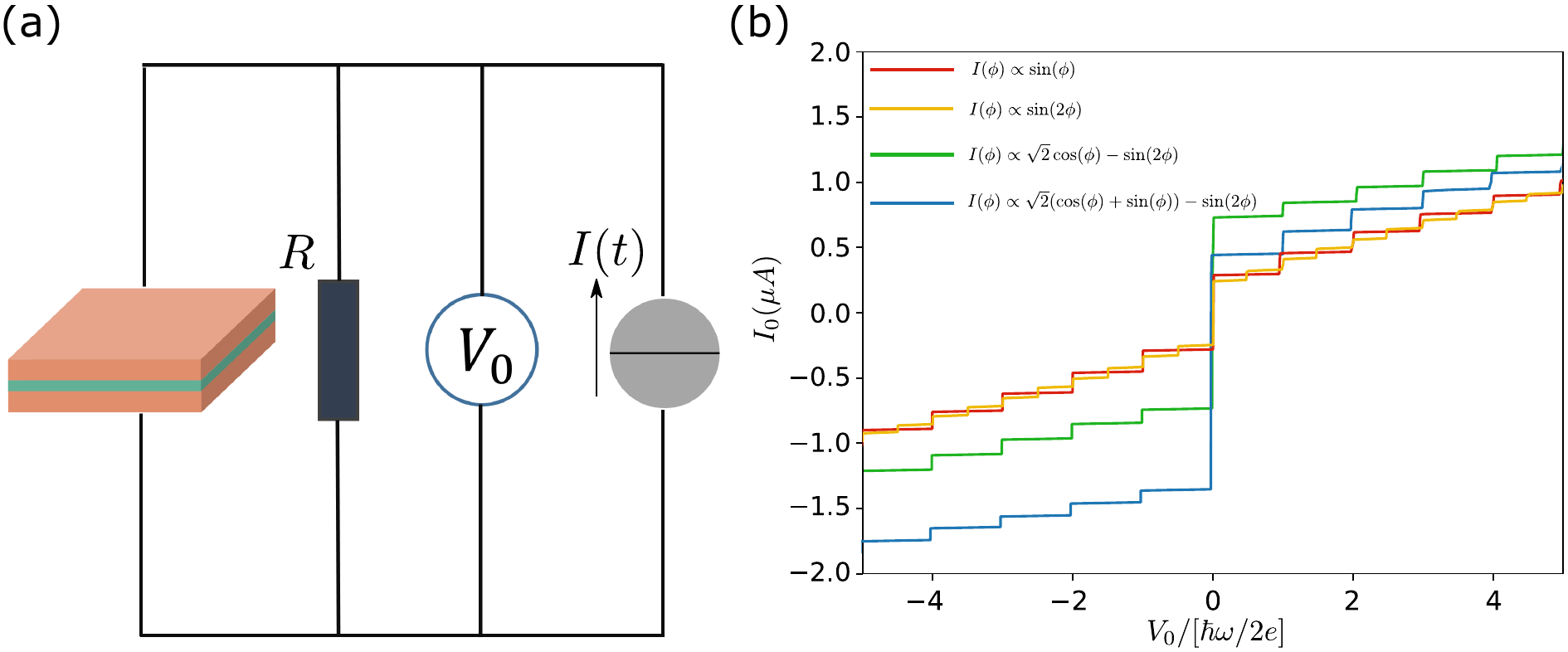}
	\caption{(a) Schematic of the Shapiro steps circuit in experiment. The RSJ model is driven by the current $I_0(t)$, and the dc voltage $V_0$ across the junction is measured. (b) Numerical result of $I_0$ changes with $V_0$ for four different current phase relation, the parameters are $E_{J,1}^{(2)}=1\mu A,I(\omega)=0.8\mu A,R=10\Omega, \omega=3.14$ GHz.
  }
\label{fig4}
\end{figure}

With the JDE and unconventional current phase relation, we then calculate the Shapiro steps for the system as an alternative
method for detecting the unconventional current-phase relation \cite{Shapiro1963,Park2021,Souto2022,Hu2023}. This calculation employs a resistively shunted Josephson junction (RSJ) model, which consists of an s-d-s Josephson junction in parallel with a resistance $R$. The current injected into the circuit contains both direct current (dc) and alternating current (ac) components: $I(t)=I_0+I_{\omega}\cos(\omega t)$. The dc voltage drop $V_0$ across the junction can be measured, as shown in Fig.~\ref{fig4}(a). In the RSJ model, the phase dynamics follows \cite{Yao2021,Park2021}
\begin{equation}
    I_0+I_{\omega}\cos(\omega t)=V/R+I(\phi),
\label{RSJ}
\end{equation}
with $V=\frac{\hbar}{2e}\dot{\phi}$ the voltage drop across the junction. The dc voltage drop $V_0$ is the time average of $V$ with $V_0=\langle V\rangle_T$. For the current phase relation $I(\phi)\propto\sin\phi$ and $\sin2\phi$, the current-voltage curve exhibits steplike behavior and remains symmetric for the voltage, $I_0(V_0)=I_0(-V_0)$, known as Shapiro steps \cite{Shapiro1963} (Fig.~\ref{fig4}(b)). The current jumps occur precisely when the average voltage matches $\langle V\rangle_T=k\hbar\omega/2e(k\hbar\omega/4e),k=0,\pm1,\pm2,\cdots$ for $I(\phi)\propto\sin\phi(\sin2\phi)$, as depicted by the red and yellow lines in Fig.~\ref{fig4}(b). In the s-d-s junction system, the current phase relation is primarily contributed by the $\cos\phi$ and $\sin2\phi$ terms if the $C_4$ rotation symmetry is present in the system. This result in a symmetric current-voltage relation, $I_0(V_0)= -I_0(-V_0)$, as shown by the green line in Fig.~\ref{fig4}(b). However, with the $C_4$ rotation symmetry breaking, the current-voltage characteristics develop an overall asymmetric nature, $I_0(V_0)\neq -I_0(-V_0)$, due to the nonreciprocal nature of the junction, as depicted by the blue line in Fig.~\ref{fig4}(b).

In experiment, the d-wave superconductors have been actively studied recently and can be gained with cuprates, such as BiSrCaCuO and YiBaCuO \cite{Mehbod1989,Topal2010,Liao2018,Lee2021,Mercado2022,Wang2023}. The implementation of JDE requires asymmetric interlayer couplings in the s-d-s junction, which can be achieved by fabricating one interface with versus without an insulating barrier at the interface in the experiment. Additionally, there are many candidates of s-wave superconductors, such as Al, Pb superconductors with critical temperature below $10$ K \cite{Ruby2015,Wang2019}, and MgB2, iron-based superconductors \cite{Nagamatsu2001,Eisterer2007,Paglione2010} with relative high critical temperatures. To lock one of the s-wave superconductor phases to its minimum point, the s-d-s Josephson junction system can also be fabricated with two s-wave superconductors having significantly different critical temperatures. Furthermore, since the Josephson coupling strength is also related to the contacting area, the s-d-s junction can also be fabricated with a large contacting area difference \cite{supp} to increase the associate Josephson coupling strength and decrease the charge energy and lock the corresponding phase to its minimum point. In this device, breaking $C_4$ symmetry is essential to generate the JDE. Besides the $C_4$ symmetry breaking in the s-d interlayer couplings due to the lattice deviates from standard square shape, it can also arise from the $C_4$ symmetry breaking in the d-wave pairing function, which is indicated in relative experimentally works \cite{Bianconi1996,Ando2002,Sato2017,Auvray2019,Zhu2021,Zhu2023}. The JDE can also exist for the $C_4$ symmetry breaking in d-wave pairing function, $\Delta_k^d=\Delta_d(\cos k_x-\cos k_y)+\Delta_{ds}$, with $\Delta_{ds}/\Delta_d$ represent the $C_4$ symmetry breaking \cite{supp}. In the experiment device, both mechanisms may coexist and the JDE can be tuned to zero by the strain \cite{supp}. Our work provides a potential experimental method to detect the presence of an s-wave pairing component in the pairing function of cuprates.

\begin{acknowledgments}
We would like to thank Noah F. Q. Yuan, Shun Wang and Cheng-Yu Yan for fruitful discussions.
\end{acknowledgments}

%

\end{document}